# A generative model of realistic brain cells with application to numerical simulation of diffusion-weighted MR signal

*Marco Palombo[1]\*, Daniel C Alexander[1], Hui Zhang[1]*

[1] Centre for Medical Image Computing and Dept of Computer Science, University College London, London, UK.

**Abstract**

In this work, we introduce a novel computational framework that we developed to use numerical simulations to investigate the complexity of brain tissue at a microscopic level with a detail never realised before. Directly inspired by the advances in computational neuroscience for modelling brain cells, we propose a generative model that enables us to simulate molecular diffusion within realistic digitalised brain cells, such as neurons and glia, in a completely controlled and flexible fashion. We validate our new approach by showing an excellent match between the morphology and simulated DW-MR signal of the generated digital model of brain cells and those of digital reconstruction of real brain cells from available open-access databases. We demonstrate the versatility and potentiality of the framework by showing a select set of examples of relevance for the DW-MR community. Further development is ongoing, which will support even more realistic conditions like dense packing of numerous 3D complex cell structures and varying cell surface permeability.

**INTRODUCTION**

Virtual histology is an emerging paradigm in medical imaging. The ultimate goal is to estimate microscopic tissue properties at the macroscopic scale using non-invasive imaging techniques, such as MRI. The current generation of non-invasive microstructure imaging techniques are rapidly becoming part of the mainstream package of imaging tools used routinely in clinical studies and exams (1-7). They primarily employ diffusion-weighted MRI (DW-MRI), which uses magnetic fields gradients to sensitize the MR image to the displacement pattern of particles, usually water molecules, within tissue (8, 9). In particular, self-diffusion of MR visible molecules in a magnetic field gradient generates a signal loss that depends on the characteristics of the gradients as well as on tissue features that hinder or restrict the diffusion process over time, like cell membranes. Thus, it can be possible to infer microscopic tissue features from the macroscopic signal loss measured by DW-MRI.

\*Corresponding author: Dr. Marco Palombo; e-mail: marco.palombo@ucl.ac.uk

Unfortunately, the relationship between the DW-MRI signal and the microstructure of complex biological tissues like the brain is still not well understood. Over the last decade, a wide variety of mathematical and biophysical models have been proposed to describe this relationship (4, 7). However, the validity of their underlying model assumptions is still under debate. In fact, the tissue microstructure is highly complex while the signal is quite simple so the mapping from signal to microstructure is highly underconstrained. This means that the more complex the system, the more challenging the identification of the most relevant features (10). Objective data-driven assessment of these assumptions, which is often hard to conduct experimentally, remains an important yet unsolved challenge.

In this context, advanced numerical simulations can provide a powerful tool to test the limits of a specific biophysical model or theory and aid the design of optimized experimental strategies. Numerical simulations and numerical phantoms play a unique role in validation that is complementary to other forms of phantoms (physical, *in vitro, ex vivo* and *in vivo*). Experiments with physical phantoms are expensive and time-consuming to set up, and often lack sufficient realism. Experiments with fixed or excised-viable tissue lack a well-defined ground truth. On the other hand, accurate and complex numerical simulations can provide a unique ground truth mapping between the relevant microstructural features and the diffusion MR signal. Although they necessarily represent a model of the real world based on our current understanding, they provide a framework for validation that is much more controlled (with known ground truth) than *in vitro*, *ex vivo* and *in vivo* phantoms, and is much more flexible than physical phantoms.

A key limitation of simulations to date is that they are too simplistic or inflexible: those based on geometric primitives are not realistic; and those derived from histological images lack flexibility. The state-of-the-art simulators for DW-MR signal are based on Monte-Carlo methods to simulate spins' diffusion within 3D digital models (substrates) representing the tissue (11, 12). These simulators are in principle able to manage complex substrates, but their usual applications have been limited to just configurations of simple geometric primitives such as cylinders and spheres (11, 13-20). Despite their simplicity, these models have been very useful to study different characteristics of brain white matter (WM) tissue. In fact, brain WM is comprised mostly of myelinated axonal bundles, and simple geometries like packed cylinders of poly-dispersed radii represent a valuable first order approximation. For instance, these kind of models informed the analysis of DW-MRI measurements in both healthy and diseased conditions, helping investigating the contribution of different WM tissue features, like axonal permeability (14, 20, 21), undulations (22), beadings (23), fiber crossing (24), and so on. Few examples exist of generating more realistic substrates for WM. Some of them use complex 3D meshes reconstructed from histological images (25). Nevertheless, these approaches do not enable users to investigate, in a controlled fashion, all the possible geometrical configurations of a complex tissue microenvironment. To address this issue, a novel algorithm to design more realistic membrane geometries better mimicking the structure of brain WM axonal bundles has been recently proposed and embedded in the DMS simulator (26). However, there is still lack of a proper computational framework that enables the creation of realistic numerical phantoms of *any* brain tissue microenvironment.

In particular, gray matter (GM) is still one of the most challenging microarchitecture to simulate. In fact, GM is comprised of complex shaped structures like brain cells (e.g. neurons and glia) densely packed together. In order to simulate a realistic substrate mimicking GM, first of all it is necessary to have a realistic digital model for the different cells in the brain. This is extremely challenging, because brain cells are complex branched structures, comprised of different connected parts, like cell body (namely soma) and cellular projections (namely neurites). Moreover, high quality meshes, ensuring correct connectivity between these distinct compartments, are often essential to obtain accurate simulations, resulting in an exponential increase of the computational complexity. To date, only a few attempts to simulate more realistic brain cell structures for DW-MRI applications have been published (27-30). However, they still rely on a simplistic description of the cell structure, as for instance 1D branched structure (29) and disconnected (27, 30) or connected cylinders (28).

Here we present new algorithms enabling for the first time the construction of ultra-realistic brain cell micro-environments and the execution of diffusion simulations within them. Specifically, we introduce a novel generative model to design realistic digital substrates of brain cells. We address two main challenges in performing numerical simulation of diffusion process within realistic 3D digital representation of brain cell:

1) Handling the large complexity of brain cell morphology, which requires a convenient digital representation that relies on a small set of controllable features, providing realism and flexibility;

2) Ensuring correct connectivity between the distinct compartments comprising the system (like each branch of a dendritic tree, cellular projections and soma, different cellular entities, etc.) minimizing the computational burden.

We validate our new approach by showing an excellent match between the morphology and simulated DW-MRI signal of the generated digital model of brain cells and those of digital reconstruction of real brain cells from available open-access databases.

**METHODS**

**Computational framework**

Here we propose a computational framework that enables us to perform numerical simulation of many particles diffusing *inside* ultra-realistic brain cellular structures.

The proposed framework is implemented in MATLAB (the Mathworks) and Python. In particular, it is designed to interface optimally with i) open-access databases of brain cell morphology, such as NEUROMORPHO (neuromorpho.org) and the Allen Brain Atlas (https://www.brain-map.org), ii) CAMINO (www.camino.org.uk) for the robust and reliable synthetic DW-MRI signal computation, and iii) standard toolboxes to visualize and analyse neuronal cell morphology like the TREES toolbox for MATLAB (www.treestoolbox.org).

A block diagram of the proposed framework is shown in **Figure 1**, where the green blocks represent the original contribution of this work: new algorithms to construct ultra-realistic digital models of brain cells. Specifically, the cell-skeleton generator and reader blocks are implemented either to generate digital model of brain cells by using a generative model (described in further details in the next section), or read digital reconstructions of real cells from experimental data or open-access databases. An efficient

3D surface mesh builder block is then designed to convert each skeletonized digital model in a 3D triangular mesh, suitable for interfacing with CAMINO or other toolboxes used in computational neuroscience, like TREES

The modular structure of the framework guarantees that the synthetic substrate generator, the diffusion simulator and the DW-MRI synthesizer operate independently, which can be an advantage due to the challenging task of creating a suitable complex geometry for spatial simulations. To compliment these possibilities, compatibility with current developed standards such as SWC, PLY and STL is also provided. Moreover, the modular structure supports straightforward parallelization of each block's tasks leading to substantial computational performance-boosts.

The framework accepts as input either a pre-built cellular skeleton, e.g. like those available on NEUROMORPHO, or an arbitrary skeleton built from scratch. A cellular skeleton generator based on an extension of the statistical model for complex cell morphology characterization introduced in (29) is implemented.

**A generative model of brain cell structure and complexity**

To model complex cell structures with correct connectivity, we describe each cellular compartment as a branched structure (*backbone*), whose individual branch has a specific radius $r_{segment}$ and each process (collection of branches sharing the same parent) radiates from the cell body (soma) of specific radius $r_{soma}$. The ensemble of backbone and sizes defines our cellular *skeleton*. The skeleton of digital cells can be either imported into the proposed framework from available public databases of real brain cell morphology or generated using the cellular skeleton generator provided. An example of 3D cell backbone and skeleton for a real Purkinje cell from NEUROMORPHO is shown in **Figure 2**.

The cell generator enables us to define realistic cellular morphology *a priori* and thus to investigate different cellular/tissue scenarios/conditions in a controlled fashion. In order to obtain realistic brain cell structures with controllable priors on the cell morphology, it is necessary to develop a generative algorithm that respects both the cell morphology priors and natural laws regulating neuronal branching. We achieve this in two steps.

First, following the paradigm introduced in (29), the overarching cellular architecture is defined by four morphometric statistics (each defined by a mean and SD) accounting for the characteristic "tree-like" structure of neurons and glia: the number of projections $N_{proj}$ leaving the soma (e.g., the dendrites or glial processes), the number of successive embranchments (bifurcations) $N_{branch}$ along each process, the segment length $L_{segment}$ (in micrometers) for a given segment of process joining two successive branching points and the bifurcation angle $\theta$. For each of these statistics, a Gaussian distribution was assumed, with SDs $SD_{Nproj}$, $SD_{Nbranch}$, $SD_{Lsegment}$, $SD_\theta$ (e.g. see **Figure 1**). This statistical model defines the basic backbone in a similar way as in (29). The backbone obtained fully respects the desired priors on the cell morphology.

Secondly, in order to improve the realism of the digital model for brain cells, the backbone is refined following the locally optimized graph approach proposed in (31). Specifically, given the set of connected points in the backbone, a distance cost function composed of two components inspired by Cajal's laws of neuronal branching is computed (31): 1) the wiring cost corresponding to the Euclidean distance to the neighboring nodes in the backbone; 2) the conduction time cost, corresponding to the path length from the soma to the point under consideration. A tunable parameter named balancing factor $b_f$ weighs these two cost functions against each other. With $b_f=0$ we have the shortest connection network, while with $b_f=1$ we get the entirely compartmentalized stellate structure, where each given point is directly connected to the soma. If the initial backbone represents the configuration that minimizes the distance cost function with the chosen $b_f$ for each node, then it is preserved. Otherwise, a minimum spanning tree algorithm (32) is used to create the final backbone, that respects both the cell morphology priors and the Cajal's laws of neuronal branching. In order to define the final skeleton, $r_{soma}$ and $r_{segment}$ can be arbitrarily defined. The cellular structure can be made more and more complex by arbitrarily defining the cell projections direct over path ratio, $\eta$, and the cell projections radius of curvature $R_c$. Dendritic spines and/or astrocytic leaflets (30) can also be added at arbitrary density $\rho_{sp}$ and size of head $h_{sp}$ and neck $n_{sp}$. The skeleton is saved in SWC format, one of the most widely used formats to store information on cellular morphology.

**Modelling cell body and branching point**

Once a cellular skeleton is provided (imported or generated), the framework generates a 3D surface mesh of the whole structure, taking into account individual branch and soma size (**Figure 2**). Because high quality meshes are essential to obtain accurate simulation results, instead of developing our own mesh generator, we make use of well validated open-source mesh generation software, specifically here BLENDER (https://www.blender.org/). We use the BLENDER "SWC Mesh" add-on (https://github.com/mcellteam/swc_mesher) to create a first fine surface mesh of the cellular skeleton (**Figure 3-a**). This procedure also ensures that the 3D surface mesh has smooth transition at critical points that connect different branches and the soma. Because of the complexity of the cellular skeleton, the surface mesh obtained usually consists of millions of faces. In order to minimize the computational burden, a home-made python script is then used to refine the mesh by progressively smoothing, reducing and triangulating the mesh to reduce the number of faces to some thousands, while keeping the overall morphology unaltered (**Figure 3-b**). Of course, it is possible to reduce the number of faces even further. However, we experimentally evaluated that some thousands of faces are a good compromise between morphology preservation and memory load. This ultimately reduces the computational complexity of the diffusion process simulation step.

The mesh generator outputs the 3D surface mesh in standard file formats, including PLY and STL, ready to be fed into CAMINO to start the numerical simulation of the diffusion process and then the corresponding DW-MRI signal computation.

**Simulating diffusion process and DW-MRI signal**

With the support for the PLY format, which is compatible with CAMINO, the framework enables DW-MRI signal synthesis with one of the most popular simulators of diffusion process and DW-MRI signal. CAMINO (33, 34) is an open-source software toolkit for diffusion MRI processing, containing a powerful and validated Monte-Carlo based molecular diffusion simulator. Briefly, the Monte-Carlo simulator engine models the population of spins as random walkers in a 3-D environment. A specific user-defined DW-MRI sequence is also modeled to track phases over the trajectories of the spins and thus derive specific DW-MRI measurements. The simulation is used to synthesize a set of noise-free measurements from diffusing spins on a specified substrate and DW-MRI sequence, to which noise can then be added. Further details on the specific algorithm used to simulate spins diffusion within a given substrate, and to compute the corresponding DW-MRI signal can be found elsewhere (11).

**RESULTS**

**Real and synthetic cells generation**

We show the potential of our new simulation framework by generating individual cells using the cell generator with different parameters, in order to obtain different cell morphologies and complexity in a controlled fashion. Specifically, we show that it is possible to simulate different brain cell types, using a set of archetypical neurons identified by Cajal (35): Purkinje cell, granule cell, motor neuron, tripolar neuron, pyramidal cell, chandelier cell, spindle neuron and stellate cell (**Figure 4-a**). The 3D generated models are reported in **Figure 4-b** and the parameters used to generate them in **Table 1** and **2**.

The same cell types reported in **Figure 4-a** can also be imported from real reconstructions available on neruomorpho.org and the corresponding 3D generated model is reported in **Figure 4-c**.

**Comparing real and synthetic cell morphologies**

We report here an example of using the proposed framework to relate detailed tissue microstructure to DW-MRI signals from the intracellular diffusing water pool only. Towards this goal, from the reconstructed real and synthetic cell structures in **Figure 4-b** and **c**, we chose three very different morphologies (Purkinje cell, motor neuron and pyramidal spiny neuron), as representative of the cell structure heterogeneity characterizing the brain tissue.

The overarching morphology of the three chosen cellular structures from real microscopy data and the corresponding synthetically generated ones are compared using dendrogram descriptors (**Figure 5-a**) and the 3D Sholl analysis (**Figure 5-b**), as provided by the TREES toolbox. Dendrograms are frequently used to illustrate the arrangement and relationship of the nodes in a graph. Here we use dendrograms to show that the overall topology of the skeleton obtained from our generative model mirrors well that of real brain cells. From **Figure 5-a** we can see that the extent, complexity and width of the dendrograms from the synthetically generated cells match very well those from the real ones. Note, however, that individual dendrograms for each pair of synthetic *versus* real cells can look slightly different, due to the randomness in the generation process, although they show the same overarching features overall. This strong match between synthetic and real cells is further confirmed by **Figure 5-b**. Indeed, the distributions of 3D Sholl metrics from the individual real and synthetically generated 3D cell structures are reported in **Figure 5-b** and were found to be not statistically significantly different (two-tailed t-test, P>0.05).

**DW-MRI features simulation**

Typically, two DW-MRI experimental strategies are used to characterize brain tissue microstructure: 1) the investigation of the dependence of the normalized signal attenuation on the b value, for a wide range of b values (4, 5); and 2) the diffusion time dependence of the apparent diffusion coefficient (ADC), for a wide range of different diffusion times (29, 36). Here we show numerical simulation results concerning these two kinds of experiments.

The three chosen complex synthetic and real cell structures are fed into CAMINO to simulate the diffusion of $5 \times 10^5$ non-interacting spins, with diffusivity $D_0=2$ $\mu m^2$/ms and $\varepsilon_t = 20$ ns. An illustrative Pulsed-Gradients-Stimulated-Echo experiment was simulated with: 30 b-values=0-30 ms/$\mu m^2$ obtained by changing only the diffusion gradient strength, 256 directions (uniformly distributed over a sphere) per b value, $\delta=3$ ms and 5 different $\Delta$ values per each set of b values: $\Delta=11, 26, 46, 76, 91$ ms. From the intracellular direction averaged DW-MRI signal at b=1 ms/$\mu m^2$, the intracellular ADC at each simulated diffusion time $t_d = \Delta-\delta/3$ was computed as ADC($t_d$) = –Ln[S(b=1, $t_d$)/$S_0$]/1 , where $S_0$ is the signal at b=0. The total computational time per cell was: ~1 hours using a single thread of a 2.4 GHz IntelCore i7; ~1.5 min parallelizing the computation on a high-performance computing cluster. This computational time should be considered only as indicative, since it depends on the complexity of the cell and the details of the mesh used, as well as the simulator setting, such as the number of spins.

**Comparing real and synthetic cell DW-MRI features**

The logarithm of the intracellular direction averaged DW-MRI signals, normalized by the signal at b=0, are reported in **Figure 5-c** as a function of b for the three illustrative cell structures chosen and $\delta/\Delta=3/11$ ms. The mean squared-difference between intracellular DW-MRI signals computed from the simulation in real 3D cell structures and from the synthetically generated ones was found to be ~ $10^{-7}$ for all the three cellular structures considered. To obtain a mean squared-difference lower than $10^{-7}$ between noise free and noisy signal, with a given finite SNR, we have estimated that SNR $\geq$ 4000 is needed. This suggests that for simulations were SNR < 4000, the two signals are indistinguishable.

In the inset in **Figure 5-c**, the diffusion time dependence of the intracellular ADCs for the three cellular structures considered are reported. The mean squared-difference between intracellular ADCs computed from the simulation in real 3D cell structures and from the synthetically generated ones was found to be ~$10^{-7}$ for all the three cellular structures considered. Following the same argument as in the previous paragraph, for simulations where SNR < 4000 the two ADC time dependences are indistinguishable.

**Effect of mesh finish on the simulated DW-MRI features**

Finally, to show the bias introduced by the mesh finish, we also performed the same Pulsed-Gradients-Stimulated-Echo experiment for the representative meshes of a Purkinje cell in **Figure 3**, using both a complex mesh of ~$10^6$ triangles (**Figure 3-a**) and an optimized minimal mesh of ~$10^4$ triangles (**Figure 3-b**).

The logarithm of the intracellular direction averaged DW-MRI signals, normalized by the signal at b=0, are reported in **Figure 3-c** as a function of b for the two illustrative meshes chosen and $\delta/\Delta=3/11$ ms. In the inset in **Figure 3-c**, the diffusion time dependence of the intracellular ADCs for the two meshes considered are also reported. The mean squared-difference between the simulated DW-MRI signals with the two meshes was found to be ~$10^{-8}$. To obtain a mean squared-difference lower than $10^{-8}$ between noise free and noisy signal, with a given finite SNR, we have estimated that SNR $\geq$ 10000 is needed. This suggests that for simulations were SNR < 10000, the two signals are indistinguishable. Same results were obtained concerning the diffusion time dependence of the ADCs computed for the two meshes. The computational time using a single thread of 2.4 GHz IntelCore i7 was ~50 hours for the finest mesh (**Figure 3-a**) and ~1 hour for the minimal mesh (**Figure 3-b**).

**DW-MRI features from selected synthetic cell types**

Finally, we show an example of current relevance for the DW-MRI scientific community. We use the computational framework to investigate whether, in ideal conditions of infinite SNR and under the experimental conditions chosen, different cell types like Purkinje cells, motor neurons, and pyramidal spiny neurons, characterized by very different morphological features, may provide different intracellular DW-MRI features.

The intracellular direction averaged signals as a function of b for the three different cell types investigated are reported in **Figure 6**. While, the intracellular ADC time dependence for the three different cell types are reported in the insets in **Figure 6**. The values of the 12 features chosen to generate the synthetic cell structures using the proposed generative model are reported as radar plots in **Figure 6** and correspond to the value reported in **Table 1** and **2** for the cell types investigated: A, C and E. We found that in the ideal case of infinite SNR and under the experimental conditions considered, the three cell types (Purkinje cells, motor neurons, or pyramidal spiny neurons) provide three different signatures in the b dependence of the intracellular direction averaged signal and in the time dependence of the intracellular ADC.

**DISCUSSION**

**The first ultra-realistic simulator of brain cell structure for DW-MRI**

The main contribution of our work is the introduction of a new generative model of brain cell morphology (green blocks in **Figure 1**). Here we demonstrated how this enables for the first time the design of numerical simulations for DW-MRI of digitalized ultra-realistic brain cell structures, achieving a new standard in fidelity. The qualitative and quantitative similarity between the structure of real brain cells reconstructed from microscopy (and public available on NEUROMORPHO) and that of the synthetically generated ones, using the generative model, is evident from **Figure 5**. The presented framework enables us to use these digital reconstructions as the basis for a potentially unlimited range of simulations, each representing an *in silico* experiment. Here we show just an example of relevance for the DW-MRI scientific community (**Figure 6**): the simulation suggests that different cell types, like Purkinje cells, motor neurons, or pyramidal spiny neurons, characterized by very different morphological features, provide different outcomes of DW-MRI experiments like time dependence of the intracellular ADC (insets, **Figure 6**) or high b value dependence of the intracellular direction-averaged signal (**Figure 6**). This suggests that it is possible to use the simulation framework to design and optimize DW-MRI protocols in order to be the most sensitive to specific features of the different cell populations in the brain, like for instance the cell body size and density (37), cell projections curvature (38), dendritic tree orientation dispersion (39), etc. In its current version, the new computational framework introduced here already provides MR scientists and engineers with a tool for exploring a large range of brain cell microstructure scenarios in a flexible and controlled fashion.

**A new generative model for controlled cell morphology modeling**

This work introduces a new paradigm for ground-truth controlled studies of brain cell structure. The generative model introduced here enables users for the first time to study in a controlled fashion the link between DW-MRI measurements and specific aspects of the brain cell morphology, like the complexity of the dendritic tree, the degree of curvature and branching of cell fibers, soma size contribution, short range disorder of synaptic buttons, spines and leaflets and much more. By describing the complex morphology of brain cells using only a small number of tunable features (12 in the current implementation), our generative model based strategy represents a unique tool for investigating selective structural alterations due to specific diseases, helping disentangling the impact of a particular disease on specific features of the cell morphology. Also, it represents the basis for novel machine learning applications, for either modelling experimental data directly, or as an intermediate step to forming specific conditional probability density functions on the cellular microstructure comprising the tissue substrate. Finally, it is also possible to expand the variety of numerical simulators, for example by integrating also powerful molecular dynamic simulators like LAMMPS (http://lammps.sandia.gov), which would make possible to realistically simulate the interaction of specific molecules within the cellular space with the cell membrane or with other molecules, for example water-water, water-proteins, water-lipid layer, metabolites-metabolites, etc.

**Faster simulation of DW-MRI features with optimized 3D mesh**

As it is easy to imagine, performing numerical simulations of the dynamics of thousands of spins in the kind of ultra-realistic cell structures shown in **Figure 2** can be extremely computationally expensive. In order to minimize the computational burden, we included in our computational framework an optimized 3D mesher that allows to reduce the number of faces comprising the 3D mesh of a single cell of a factor $10^2$-$10^3$, without compromising the performance of the simulation. In **Figure 3**, we provide an example of two meshes, one comprised of ~$10^6$ triangular faces (**Figure 3-a**) and an optimized one, comprised of only ~$10^4$ triangular faces (**Figure 3-b**). The overall morphology of the cellular mesh is perfectly preserved, despite its complexity is reduced of a factor $10^2$. **Figure 3-c** shows that the simulated DW-MRI features computed in the two meshes are practically indistinguishable, but the use of the less complex mesh reduced the computational time on a single computational core by a factor

~50. Optimized 3D meshing together with parallelization make possible simulations of up to thousands of different ultra-realistic cells in less than 5 minutes (these estimates are of course indicative, since the exact computational time depends on many factors like the specific characteristics of the computing cluster used, the complexity of the cell and the details of the mesh used, as well as the simulator setting, such as the number of spins).

**Potential applications**

The computational framework introduced here represents a unique tool to test the validity of basic assumptions in current biophysical models used to estimate specific brain microstructural features such as neurites density and dispersion. Indeed, the full power of our new computational framework lies in hypothesis testing and experimental design. For example, to model the intracellular direction averaged DW-MRI signal in GM, often the dendritic tree of brain cells is modelled as independent randomly oriented sticks (40-42) or cylinders (27, 43). In reality, it is comprised of long curved and branching fibers. The computational framework introduced here can help validating the underlying hypothesis of these models showing whether or not, under specific experimental conditions, the effect of cell fibers branching and curvature is negligible. The ground-truth realistic digitalised models of brain cells introduced here can be hereafter devised for investigating many experimental questions that remain mostly unanswered, like (10): what causes the observed time dependence of intracellular biological water diffusivity along the fibers or in the gray matter? Is it varicosities, beads, synaptic boutons, undulations, or something else? Which of these structural units' changes in pathology could be detectable?

The purpose of the present work is to introduce the new computational framework and provide proof-of-concept applications in DW-MRI. Thus, the usage of the computational framework for specific hypothesis testing and experimental design will be subject of future works, including direct comparison with experimental data.

**Limitations**

The effects of $T_1/T_2$ relaxation have not been taken into account for all the performed simulations, although they clearly influence the SNR when long echo times are chosen. Also, although it is possible with CAMINO to simulate the actual level of noise corrupting the signal of real acquisitions, in this work we did not perform any SNR study since the aim of the work is not to study how different tissue microarchitecture impact the DW-MRI signal, but it is to provide a proof-of-concept of the several potentialities offered by the novel simulation framework proposed. The study of how different cellular or tissue microarchitectures impact the DW-MRI signal with realistic SNR can be the topic of future studies where the new simulation framework can be used and exploited at its full potential.

Moreover, for similar reason, cell membrane permeability has not been considered in the simulations reported here. However, it is important to underline that it is possible to include cell membrane permeability in the simulation of the diffusion dynamics, using CAMINO. Furthermore, it is also possible to assign different cell membrane permeability to different cell types or cell subparts within the same substrate. This can be useful for studies aiming at investigating the effect of brain cell membrane permeability on water diffusion in a more realistic way, since it has been shown that, for instance, the membrane of glial cells and neurons have different permeability properties (44).

One more limitation of the proposed computational framework, in its current implementation, is that it can be used to study many different brain cells, but only considering them as independent and non-interacting part of the whole brain tissue. How we are planning to extend the current framework to multiple packed cells, including extra-cellular space, is addressed in the following section. Here, we would like to underline how, even in its currently limited implementation, the proposed computational framework already represents an incredibly valuable tool for the MRI community. It is the very first framework for ultra-realistic cellular structure simulation, designed specifically for the MRI community. As such, it represents a first step, from which starting a collaborative effort to push further the current limitations of numerical simulations, and open the way towards a unique alternative to expensive physical phantoms and invasive/destructive histology sampling.

**Future perspectives: towards an ultra-realistic simulator of the brain tissue for DW-MRI**

The morphologies of the brain cells are highly diverse and variant (see **Figure 4**). The variance, presumably, originates also from their constraining arrangement in a densely packed brain substrate. Thus, in order to generate a digital representation of the brain tissue realistically, it is necessary to develop an efficient context-aware cell packing algorithm. The proposed computational framework is already designed in order to accommodate a "tissue generator" module (see **Figure 1**) that can include different algorithms for context-aware packing of the digital cells, either generated or imported from real data. This represents the predominant direction of our future work.

**CONCLUSION**

In this work, we introduce a new framework that we developed to use numerical simulations to investigate the complexity of brain tissue at a microscopic level with a detail never realized before. Directly inspired by the advances in computational neuroscience for modelling brain cells, the generative model proposed here enables for the first time numerical simulation of molecular diffusion within ultra-realistic, controlled and flexible digital brain cells, such as neurons and glia. Here we demonstrate the versatility and potentiality of the framework by showing a select set of examples of relevance for the DW-MR community. Further development is ongoing, which will support even more realistic conditions like dense packing of numerous 3D complex cell structures and varying cell surface permeability.

**Acknowledgment**

This work was supported by EPSRC (EP/G007748, EP/I027084/01, EP/L022680/1, EP/M020533/1, N018702).

# Tables

**Table 1.** Set of tunable parameters describing the morphology of the 3D cellular structures in **Figure 4b**. $N_{proj}$: number of cellular projections radiating from the soma; $N_{branch}$: number of consecutive bifurcations; $L_{segment}$: length of each individual segment comprising the cellular projections (in µm); $\theta$: bifurcation angle (in radiants); $R_c$: radius of curvature of individual segments (in µm); $\eta$: direct over path ratio of individual segments; $b_f$: balancing factor in the extension to the minimum spanning tree algorithm.

| Cell type | $N_{proj}$ (mean ± SD) | $N_{branch}$ (mean ± SD) | $L_{segment}$ (mean ± SD) | $\theta$ (mean ± SD) | $R_c$ (mean ± SD) | $\eta$ (mean ± SD) | $b_f$ |
|---|---|---|---|---|---|---|---|
| A | 1 ± 0  | 8 ± 2 | 20 ± 5  | $\pi/6 \pm \pi/8$   | 500 ± 100 | 0.80 ± 0.1 | 0.5 |
| B | 2 ± 0  | 3 ± 1 | 20 ± 10 | $\pi/8 \pm \pi/16$  | 100 ± 20  | 0.70 ± 0.1 | 0.5 |
| C | 20 ± 5 | 3 ± 1 | 40 ± 10 | $\pi/4 \pm \pi/16$  | 500 ± 100 | 0.85 ± 0.1 | 0.5 |
| D | 2 ± 0  | 3 ± 1 | 30 ± 10 | $\pi/4 \pm \pi/8$   | 500 ± 100 | 0.85 ± 0.1 | 0.5 |
| E | 10 ± 5 | 4 ± 1 | 30 ± 5  | $\pi/8 \pm \pi/8$   | 500 ± 100 | 0.95 ± 0.1 | 0.5 |
| F | 10 ± 5 | 3 ± 1 | 40 ± 10 | $\pi/4 \pm \pi/16$  | 100 ± 20  | 0.70 ± 0.1 | 0.5 |
| G | 2 ± 0  | 2 ± 1 | 30 ± 15 | $\pi/16 \pm \pi/32$ | 100 ± 20  | 0.90 ± 0.1 | 0.5 |
| H | 8 ± 2  | 3 ± 1 | 40 ± 10 | $\pi/4 \pm \pi/16$  | 50 ± 10   | 0.60 ± 0.1 | 0.5 |

**Table 2.** Set of tunable parameters describing the sizes of the 3D cellular structures in **Figure 4b**. $r_{soma}$: radius of the cell body, namely soma (in µm); $r_{segment}$: radius of each individual segment comprising the cellular projections (in µm); $\rho_{sp}$: density of spines/leaflets (in µm$^{-1}$); $h_{sp}$: radius and length of the spines/leaflets head (in µm); $n_{sp}$: radius and length of the spines/leaflets neck (in µm).

| Cell type | $r_{soma}$ | $r_{segment}$ (mean ± SD) | $\rho_{sp}$ (mean ± SD) | $h_{sp}$ (mean ± SD) | $n_{sp}$ (mean ± SD) |
|---|---|---|---|---|---|
| A | 25 | 1 ± 0.5   | 0 ± 0   | n. a.       | n. a.        |
| B | 8  | 0.5 ± 0.1 | 0 ± 0   | n. a.       | n. a.        |
| C | 10 | 0.5 ± 0.1 | 0 ± 0   | n. a.       | n. a.        |
| D | 8  | 0.5 ± 0.1 | 0 ± 0   | n. a.       | n. a.        |
| E | 25 | 0.5 ± 0.1 | 20 ± 10 | 0.5 ± 0.25  | 0.12 ± 0.06  |
| F | 5  | 0.5 ± 0.1 | 0 ± 0   | n. a.       | n. a.        |
| G | 12 | 0.5 ± 0.1 | 0 ± 0   | n. a.       | n. a.        |
| H | 10 | 0.5 ± 0.1 | 0 ± 0   | n. a.       | n. a.        |

# Figures captions

**Figure 1.** (*Center*) Schematic of the proposed computational framework. The toolbox we developed is comprised of the green blocks. It accepts as input either a pre-built cellular skeleton, like those available on NEUROMORPHO (neuromorpho.org), or an arbitrary skeleton built from scratch using a novel generative model. (*Left*) We show some example of using the digital cell generator to generate cell structures of increasing complexity in a controlled fashion from morphometric statistics *priors*: different values of mean and standard deviation of $L_{segment}$, $N_{branch}$ and $N_{process}$, and fixed values of $\theta$, radius of branches and soma. (*Right*) The toolbox outputs a 3D mesh of the whole digital cell (in the figure the scale-bars refer to 100 µm). This 3D is saved in suitable format for interfacing with CAMINO or other toolboxes used in computational neuroscience, like TREES.

**Figure 2.** Example of 3D cell backbone, skeleton and mesh for a reconstructed Purkinje cell from the NEUROMORPHO database (neuromorpho.org)

**Figure 3.** *a)* complete triangular mesh (~ $10^6$ triangular faces) obtained from the 3D mesher block from the reconstructed Purkinje cell reported in **Figure 2**. *b)* reduced mesh (~ $10^4$ triangular mesh) after progressive smoothing, reduction and triangulation of the mesh more complex mesh in a). *c)* Direction averaged normalized DW-MR signal as a function of b for the two meshes in a) and b), as computed from the simulation of $5 \times 10^5$ non-interacting spins, with diffusivity $D_0=2$ µm$^2$/ms, and a PGSE sequence with 30 b-values=0-30 ms/µm$^2$ obtained by changing only the diffusion gradient strength, 256 directions (uniformly distributed over a sphere) per b value, $\delta/\Delta=3/11$ ms. *c) - inset*, the ADC as function of diffusion time as computed from the simulation of $5 \times 10^5$ non-interacting spins, with diffusivity $D_0=2$ µm$^2$/ms, and a PGSE sequence with b=1 ms/µm$^2$, 256 directions (uniformly distributed over a sphere), $\delta=3$ ms and 5 different $\Delta$ values per each set of b values: $\Delta=11, 26, 46, 76, 91$ ms.

**Figure 4.** *a)* Set of archetypical neurons identified by Cajal (35): (A) Purkinje cell, (B) granule cell, (C) motor neuron, (D) tripolar neuron, (E) pyramidal cell, (F) chandelier cell, (G) spindle neuron and (H) stellate cell. *b)* 3D digital model of the same cell types in a) as generated using the generative model with the parameters in **Table 1** and **2**. *c)* 3D reconstruction from histological data of the same cell types in a) from open-source database neuromorpho.org.

**Figure 5.** *a)* Comparison of the dendrograms of the morphology of three selected cell types (Purkinje cells, motor neuron and spiny pyramidal neuron) obtained from 3D reconstruction from histological data of real cells (named Real) and the digital model obtained using the generative model (named Synthetic). *b)* Comparison of the probability density distributions of 3D Sholl metrics for the real and synthetic cells. *c)* Comparison of the direction averaged normalized DW-MR signal as a function of b for the three real and synthetic cells, as computed from the simulation of $5 \times 10^5$ non-interacting spins, with diffusivity $D_0=2$ µm$^2$/ms, and a PGSE sequence with 30 b-values=0-30 ms/µm$^2$ obtained by changing only the diffusion gradient strength, 256 directions (uniformly distributed over a sphere) per b value, $\delta/\Delta=3/11$ ms. *c) - inset*, Comparison of the ADC as function of diffusion time as computed from the simulation of $5 \times 10^5$ non-interacting spins, with diffusivity $D_0=2$ µm$^2$/ms, and a PGSE sequence with b=1 ms/µm$^2$, 256 directions (uniformly distributed over a sphere), $\delta=3$ ms and 5 different $\Delta$ values per each set of b values: $\Delta=11, 26, 46, 76, 91$ ms.

**Figure 6.** Example of possible application of the proposed simulation framework to investigate the DW-MR signal features originating from three very different digital cell models (corresponding to Purkinje cells, motor neuron and spiny pyramidal neuron) obtained using the controlled and flexible generative model using the parameters value reported in the radar plots and in **Table 1** and **2**. Direction averaged DW-MR signal as a function of b and ADC as function of diffusion time (inset) for the three digital model of brain cells, as computed from the simulation of $5 \times 10^5$ non-interacting spins, with diffusivity $D_0=2$ µm$^2$/ms, and a PGSE sequence with 30 b-values=0-30 ms/µm$^2$ obtained by changing only the diffusion gradient strength, 256 directions (uniformly distributed over a sphere) per b value, $\delta=3$ ms and 5 different $\Delta$ values per each set of b values: $\Delta=11, 26, 46, 76, 91$ ms.

# Figures

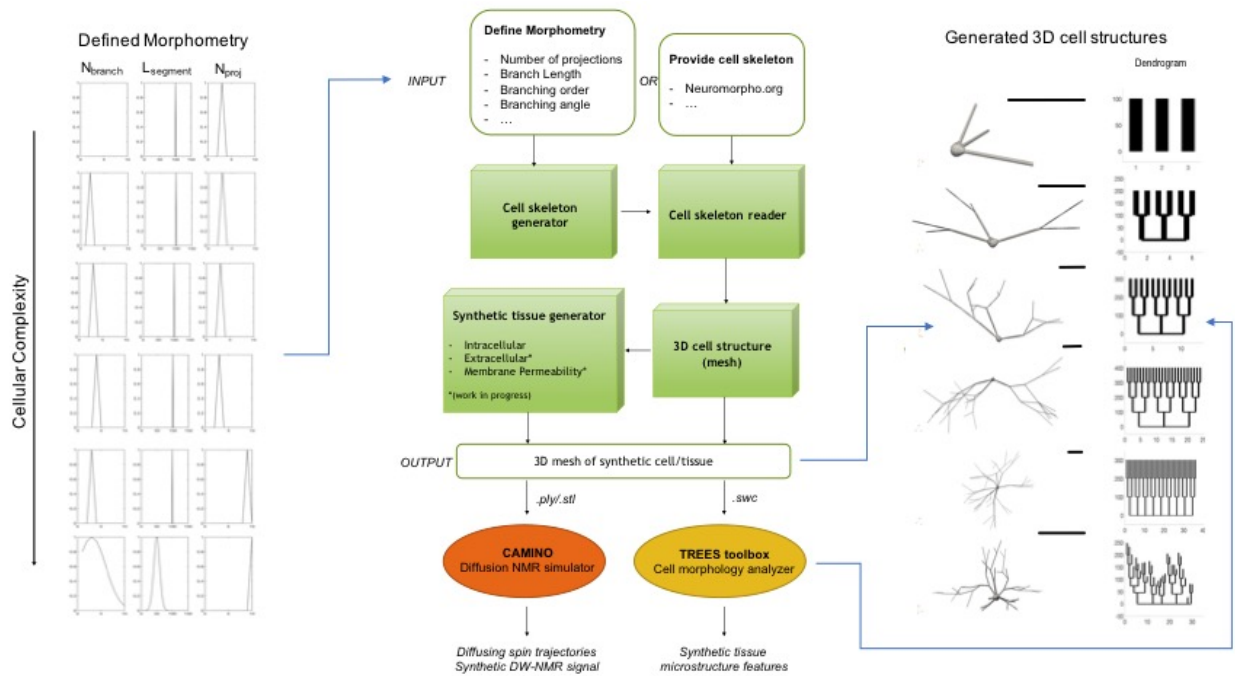

**Figure 1.**

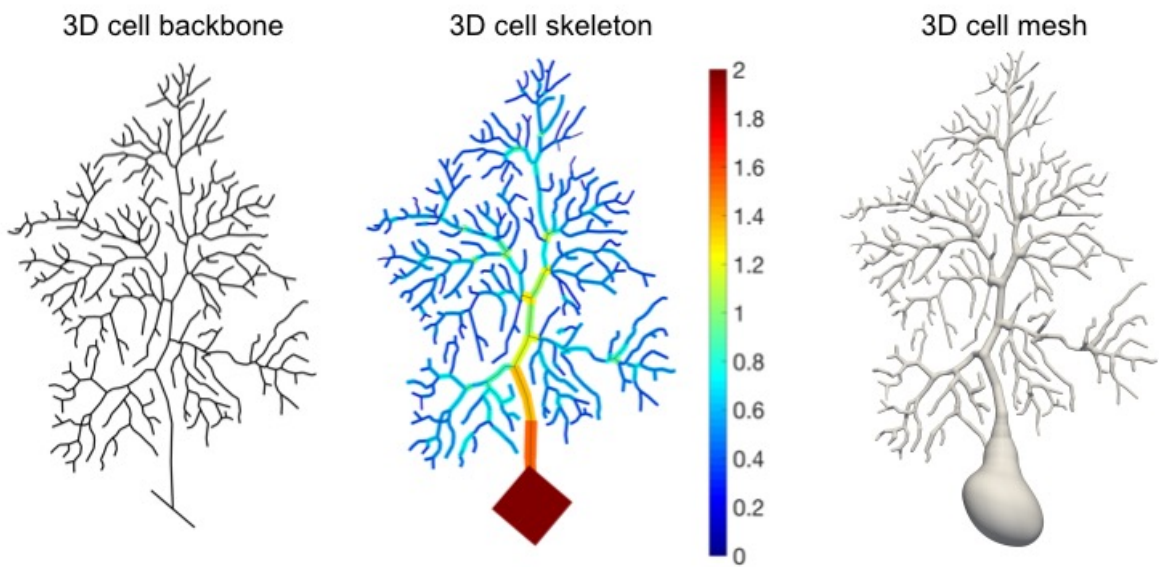

**Figure 2.**

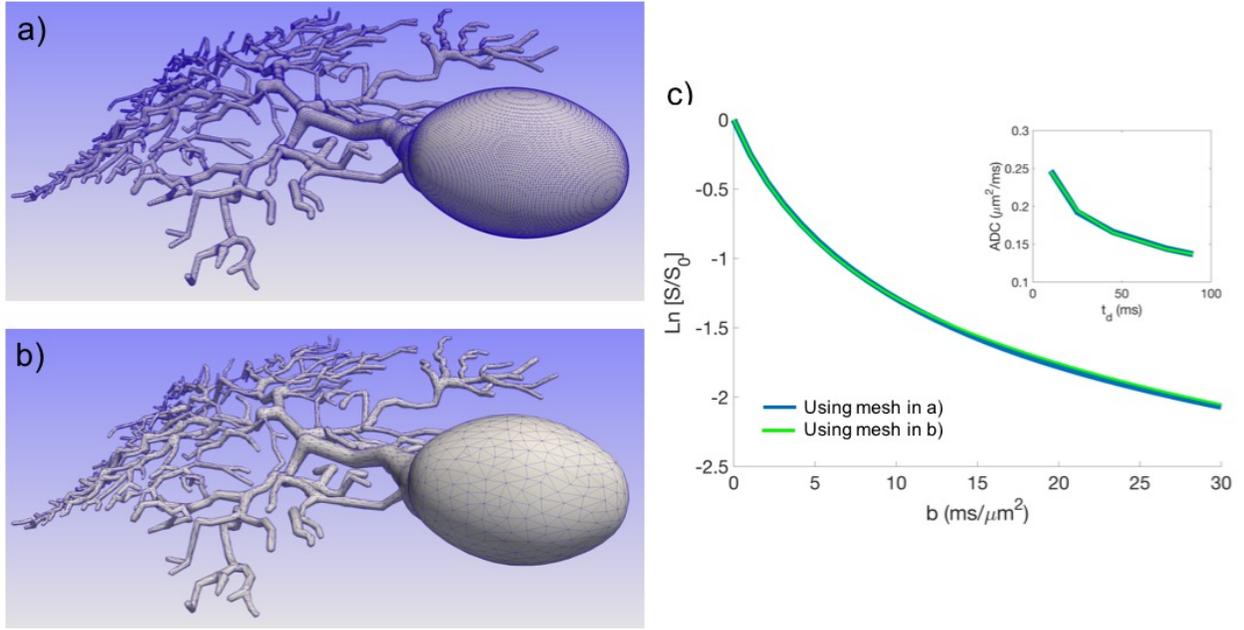

**Figure 3.**

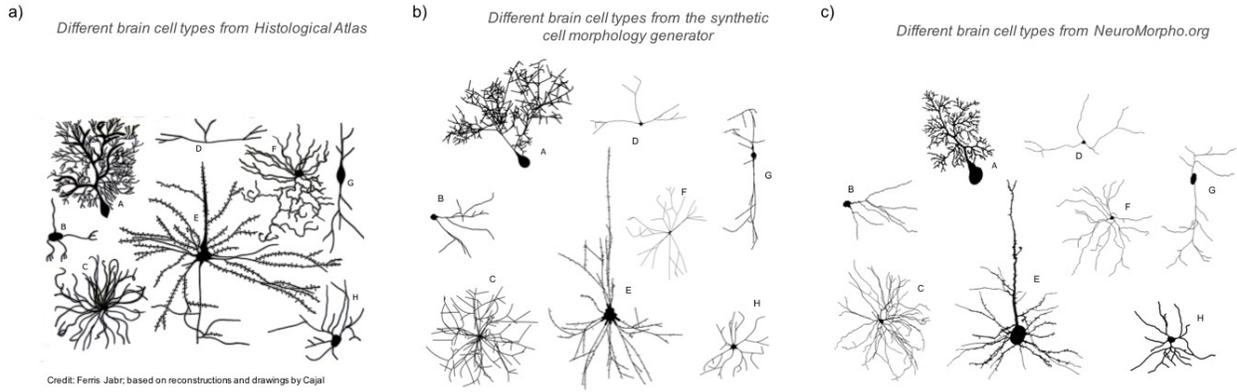

**Figure 4.**

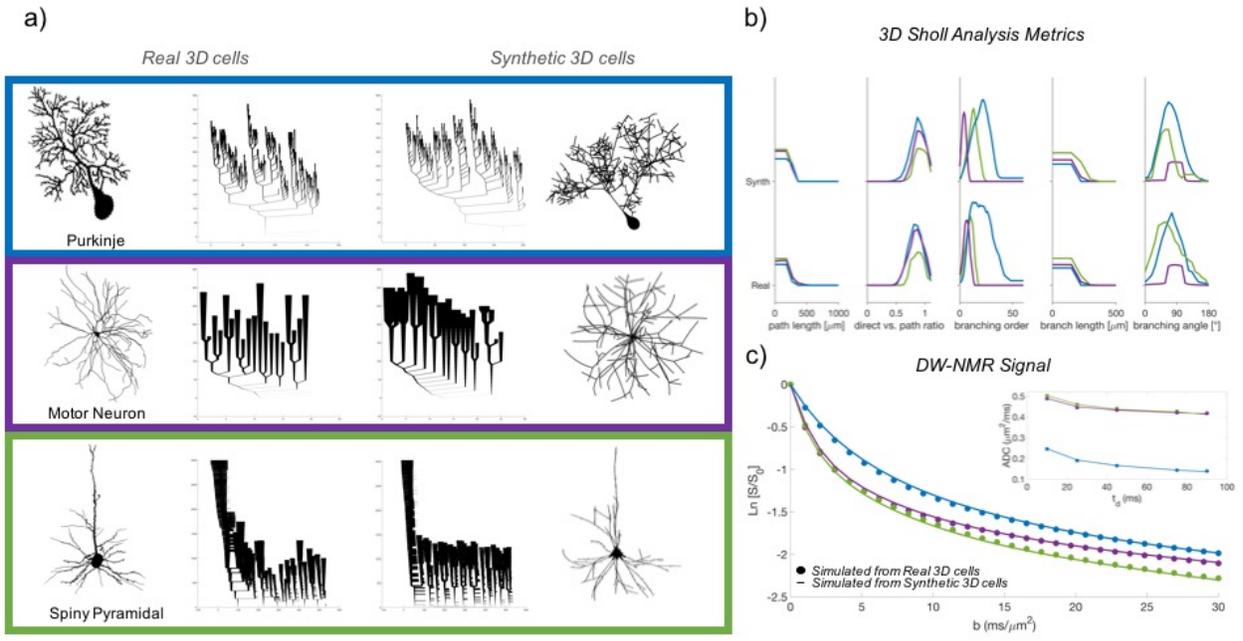

**Figure 5.**

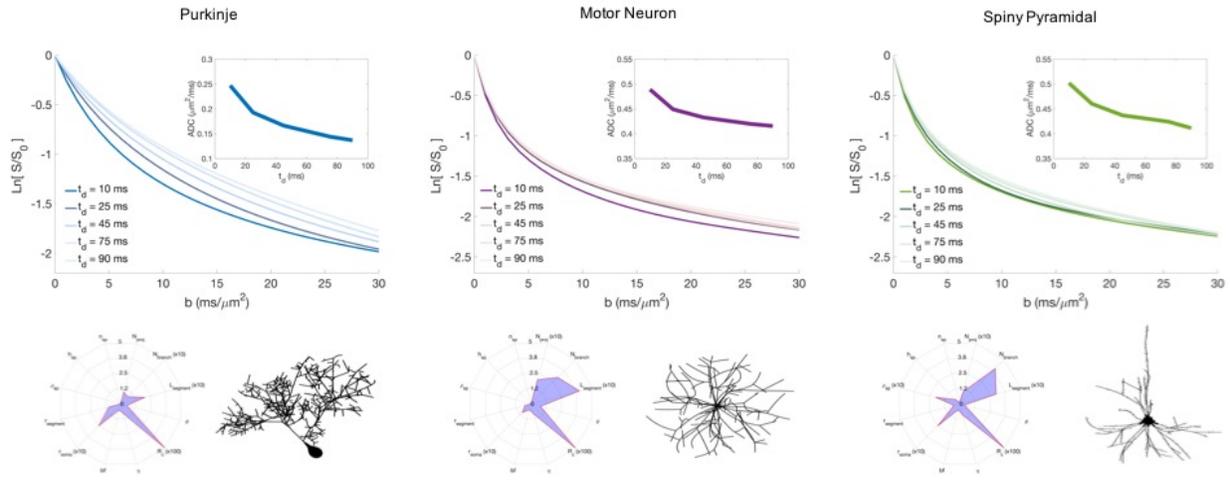

**Figure 6.**